\documentclass[doublecol,figures]{epl2}
\usepackage[latin1]{inputenc} 
\usepackage{amsmath}
\usepackage[english]{babel}
\newcommand{\ie}{i.~e.,} 
\newcommand{\eg}{e.~g.,}

\newcommand{\hc}{{\rm{H.\ c.}}}  
\newcommand{\ket}[1]{|#1\rangle}

\newcommand{\matel}[3]{\langle#1|#2|#3\rangle}

\newcommand{\gc}{{\mathcal G}_2} 
\newcommand{\gu}{G_{SET}^S}
 \newcommand{\gs}{{G_{s}}}
 
\newcommand{\cd}{c_{d}} 
\newcommand{\hd}{H_d}
\newcommand{\nd}{n_{d}}
\newcommand{\ndu}{n_{d\uparrow}}
\newcommand{\ndd}{n_{d\downarrow}}
\newcommand{\ed}{\varepsilon_d}
\newcommand{\nq}{{{\mathcal N}_q}}

\newcommand{\fone}{f_1} 
 
\newcommand{\dq}{d_q}

\newcommand{\onlinecite}[1]{\cite{#1}}

\newcommand{\rhoq}{\rho_{q}(\esc, \tsc)}

\newcommand{\esc}{\mathcal{E}}
\newcommand{\tsc}{\mathcal{T}}
\newcommand{\dd}{\overline{d}}
\newcommand{\ddd}{{\overline{d}}^{\,\dagger}}
\newcommand{\ha}{H_A}
\newcommand{\heta}{H_\eta}
\title{Thermal dependence of the zero-bias conductance through a
  nanostructure} 
\author{
  A. C. Seridonio\inst{1,2} 
  \and M. Yoshida\inst{3} 
  \and L. N. Oliveira \inst{1}
} 
\shortauthor{Seridonio, Yoshida, and Oliveira}
\institute{
  \inst{1}Departamento de F\'{\i}sica e Inform\'{a}tica, Instituto
  de F\'{\i}sica de S\~{a}o Carlos, Universidade de S\~{a}o Paulo,
  369, S\~{a}o Carlos, SP, Brazil\\
  \inst{2}ICCMP - International Center for Condensed Matter Physics,
  Universidade de Bras\'{\i}lia, 04513, Bras\'{\i}lia, DF, Brazil\\
  \inst{3}Departamento de F\'{\i}sica, Instituto de Geoci\^{e}ncias
  e Ci\^encias Exatas, Universidade Estadual Paulista, 13500, Rio
  Claro, SP, Brazil}
\date{\today{}}


\abstract{
We show that the conductance of a quantum wire side-coupled to a
quantum dot, with a gate potential favoring the formation of a dot
magnetic moment, is a universal function of the temperature. Universality
prevails even if the currents through the dot and the wire interfere.
We apply this result to the experimental data of Sato et
al.\ [Phys. Rev. Lett. {\bf 95}, 066801 (2005)].
}

\begin{document}
\maketitle

Nanodevices owe much of their development to the theory of many-body
phenomena. Consider, e.~g., the {\em single electron transistor}
(SET), a quantum dot bridging two otherwise independent
two-dimensional electron gases \cite{GSM+98.156,GGK+98:5225}. The
competition between the Coulomb blockade, which bars transport through
the dot, and the Kondo screening of the dot magnetic moment by the
electron gases \cite{hewson93}, which favors low-temperature
conduction, was discussed on blackboards \cite{GR87:452} a decade
before it surfaced in the laboratory \cite{GSM+98.156}. By the time
the first device was developed, quantitatively accurate theoretical
results were available. Chiefly important was the universal
conductance curve $\gu(T)$ for the symmetric Anderson model
\cite{CHZ94.19,BCP08:395}, which was shown to match the temperature
dependence of the zero-bias conductances in SETs and analogous
devices.

\begin{figure}
\onefigure[clip=true,width=\columnwidth]{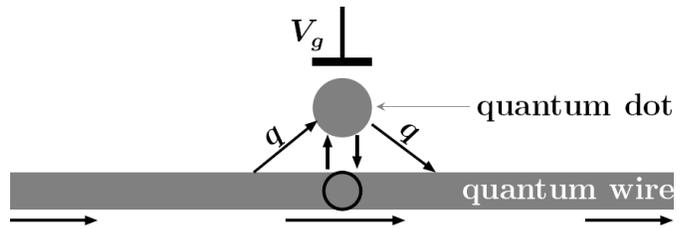}
  \caption{Side-coupled device. The gate potential $V_g$ controls the
  energy $\ed$ of the quantum-dot level $\cd$. The open circle depicts
  the Wannier orbital $f_0$.\label{fig:1}}
\end{figure}

More recently, experiment has leaped ahead of theory. The development
of complex structures, such as the {\em side-coupled} device
\cite{KAS+04:035319,SAK+05:066801,KSA+06:36,%
  AEO+06:195329,FBI+06:205326,OAK+07:084706}, has motivated only
qualitative predictions.  As Fig.~\ref{fig:1} shows, the current in
the side-coupled device is carried by electrons that can either
traverse the quantum wire or hop to a quantum dot to skip the central
section \cite{OAK+07:084706}. A Fano parameter $q$ \cite{Fa61:1866},
defined below, measures the amplitude for the latter process relative
to that for the former. The limit $q\to\infty$ emulates a SET. For
smaller $q$'s, the wire bypasses the Coulomb blockade and allows
high-temperature conduction. Below the Kondo temperature $T_K$, the
screening of the magnetic moment enhances the electronic flux through
the dot and allows interference with the flow along the central
portion of the wire \cite{HKS01:156803,FFA03:155301}.

Attentive to the diversity of experimental findings, we have applied
numerical renormalization-group (NRG) tools to an Anderson Hamiltonian
modeling the side-coupled device. The resulting essentially
exact numerical data for the temperature-dependent conductance
$G_q(T)$ will be detailed elsewhere \cite{smo08}. Here, we focus the
relation between $G_q(T)$ and the universal curve $\gu(T)$. Our
central result covers the Kondo domain, the set of dot energies and
dot-wire couplings that favor the formation of a Kondo screening
cloud.

Since the thermodynamical properties of the Kondo crossover are
universal functions of the temperature scaled by $T_K$
\cite{Wi75:773,KWW80:1003,OW81.1553,TW83:453,LWC+87:1232,SLO+96:275,%
  GGK+98:5225,WFF+00:2105,CLGp02:226805,LSB02:725}, a mathematical
relation between $G_q(T/T_K)$ and the universal function $\gu(T/T_K)$
is hardly surprising. Nonetheless, the diversity manifest in
conductances that, in contrast with $\gu(T)$, rise with temperature
and in conductance profiles (fixed-$T$
conductance-vs.\ gate-voltage plots) that display antiresonances
\cite{OAK+07:084706} rules out a proportionality between $G_q(T)$ and
$\gu(T)$. Instead, we will show that a linear mapping binds the two
functions:

\begin{equation}
  \label{eq:2}
  G_q(\frac{T}{T_K}) - \frac{\gc}2 =
  \left(\gu(\frac{T}{T_K}) -\frac{\gc}2\right)\cos2\delta,
\end{equation}
where $\delta$ is the ground-state phase shift of the wire electrons, and
$\gc\equiv2e^2/h$, the quantum conductance.

While the mapping~(\ref{eq:2}) is universal, the phase shift and Kondo
temperature are model-parameter dependent. At fixed temperature, the
Fano parameter $q$ controls the functional dependence of the
conductance on the gate voltage $V_g$. As $q$ grows, valleys in the
conductance profiles evolve into plateaus, a result in qualitative
agreement with measurements. Most importantly, Eq.~(\ref{eq:2})
affords quantitative comparison with experiment; as an illustration,
we will present curves that reproduce the temperature-dependent
conductances reported by Sato et al.\ \cite{SAK+05:066801}; show that
the dot moment was fully screened; and extract $T_K$ and $\delta$ from
the data. Our results justify mathematically the authors'
phenomenological treatment of their results.

{\em Overview.} Preliminary to the formalism, we present an overview of
conduction in the side-coupled device. We consider weak coupling to
the wire, so that the dot occupation $n_{d}$ is nearly conserved,
and illustrate the discussion with NRG plots of the
temperature-dependent conductance.

\begin{figure}
\onefigure[width=1.0\columnwidth]{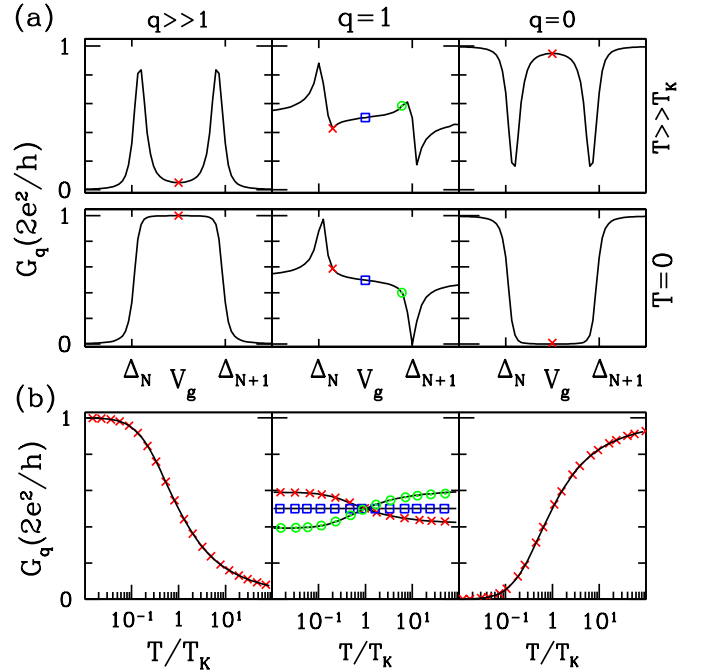}
\caption{Bird's eye view of conduction through the side-coupled
  device.  (a) Condutance $G$ as a function of applied gate voltage
  for three representative Fano parameters $q$ at temperatures high
  (top panels) or low in comparison with the Kondo temperature
  $T_K$. (b) NRG results for $G(T)$. The crosses, squares,
  and circles in each panel represent conductances
  calculated at the voltages indicated by the same symbol in the
  panels directly above it. The solid lines depict Eq.~(\ref{eq:2}),
  with $\delta$ extracted from the low-energy fixed-point eigenvalues
  \cite{KWW80:1044}.\label{fig:2}}
\end{figure}

The device has three characteristic energy scales, set by (i) the
coupling to the wire, which broadens the dot levels; (ii) the
electrostatic barrier $\Delta_{N}$ between adjacent dot occupations
$n_d=N-1$ and $n_d=N$; and (iii) the Kondo temperature $T_K$,
below which the wire electrons screen the dot moment. The first two
scales catch the eye in conductance profiles. The third one defines
the thermal regimes $T\gg T_K$ and $T\ll T_K$ displayed schematically
in the top and bottom panels of Fig.~\ref{fig:2}a, respectively.

\noindent{\em Left: $q\gg1$.} If $T\gg T_K$ (top), the Coulomb blockade impedes
conduction; to defeat it, the gate voltage must be raised, so that
the $n_d=N$ and the $n_d=N+1$ ground states are nearly degenerate. The
conductance profile is hence a sequence of narrow resonances. Upon cooling
(bottom), little changes if $V_g$ makes $n_d$ even.  For
odd $n_d$, however, the Kondo hybridization
between the wire and the dot states allows conduction. The conductance
$G(V_g)$ alternates between insulating valleys ($G=0$ for even
$n_d$) and {\em Kondo plateaus} ($G=\gc$ for odd
$n_d$) \cite{WM94:11040}.

\noindent{\em Right: $q=0$.} The pattern is reversed. If $T\gg T_K$ (top),
except at the resonant voltages, the flux through the wire is
ballistic. On resonance, the strong coupling to the dot blocks
conduction, and $G(V_g)$ dips to zero \cite{SSI+06_096603}. For $T\ll
T_K$ (bottom) and odd $n_d$, the screening cloud blocks the
wire. The conductance thus alternates between ballistic plateaus and
{\em Kondo valleys} \cite{HKS01:156803}.

\noindent{\em Center: $q=1$.} There are now two conduction paths. If $T\gg
T_K$ (top), Fano antiresonances near the resonant voltages signal
interference between the two currents. Off resonance, the
electrons flow only through the wire, and the conductance
remains close to $\gc/2$.  For $T\ll T_K$ (bottom) and odd $n_d$,
again the Kondo hybridization to the wire allows
conduction through the dot. The Kondo cloud nonetheless blocks
conduction through the wire, so that the off-resonance conductance is
again close to $\gc/2$.

For each voltage identified by two crosses, squares, or circles in
Fig.~\ref{fig:2}a, Fig.~\ref{fig:2}b displays our NRG results for
the conductance as a function of $T$ normalized by $T_K$\textemdash the
temperature at which the conductance is half the quantum conductance,
$G_q(T=T_K)\equiv\gc/2$.  The agreement with the solid lines
representing Eq.~(\ref{eq:2}) is very good: even at the limits of the
Kondo regime (crosses and circles, $T=50\,T_K$) the absolute
deviations are smaller than $0.01\,e^2/h$.

For $q\gg1$ (left, $\delta\approx0$), the device mimics a SET, the
mapping~(\ref{eq:2}) reduces to an identity, and the conductance
decays from ballistic to zero along the universal curve $\gu(T)$. The
opposite extreme, $q=0$ (right, $\delta\approx\pi/2$), reverses the
pattern: ballistic conductance at $T\gg T_K$, and perfect insulation
at $T=0$. As intuition would dictate, and as general
\cite{SAK+05:066801} and mathematical \cite{MNU04:3239} arguments
suggest, the temperature dependence complements the universal
function: $G_q(T) = \gc -\gu(T)$.  For $q\sim1$ (center,
$\delta\approx\pi/4$), three voltages $V_g$ are highlighted. In each
case, the conductance interpolates monotonically the high- to the
low-temperature limits in Fig.~\ref{fig:2}a. For $2\ed+U=0$, in
particular, Fig.~\ref{fig:2}a shows that $G_q(T\gg
T_K)=G_q(T=0)=\gc/2$; the conductance must therefore be constant,
$G_q(T)=\gc/2$, and this is what Eq.~(\ref{eq:2}) predicts for
$\delta=\pi/4$.

{\em Model.} Hamiltonians describing side-coupled nanostructures, in
the geometry of Fig.~\ref{fig:1}
\cite{HKS01:156803,THC+02:085302,MNU04:3239,SSI+06_096603,OAK+07:084706},
or other arrangements \cite{ZB06.035332,BTD06:205301,SSS09:095003}
have appeared in print. While our analysis could start from any of the
former, to keep the presentation self-contained we define the
alternative three-component Hamiltonian $H=H_d+H_w+H_{wd}$. Here, $\hd
\equiv {\ed}\nd + U \ndu\ndd$, models the quantum-dot, where
$\nd\equiv \cd^\dagger \cd$ is the dot-level occupation; $\ed$ is the
dot energy, controlled by the potential $V_g$; and $U$ is the Coulomb
repulsion. The second term, $H_w\equiv\sum_k\epsilon_kc_k^\dagger
c_k+(K/N)\sum_{kk'}c_k^\dagger c_{k'}$, models the wire, of length
$L$. The energies $\epsilon_k$, measured from the Fermi level, form a
symmetric, structureless, half-filled conduction band of width $2D$
comprising $N$ levels separated by the splitting $\Delta\equiv 2D/N$
\cite{even}. The scattering $K$ describes the (fixed) gate potential
applied to the wire \cite{OAK+07:084706}.

The last component, $H_{wd}\equiv\sum_k (V_k/\sqrt{N})c_k^\dagger
\cd+\hc$, couples the wire to the dot. The relatively small thermal
energies $k_B T\ll D$ authorize retention of the two leading terms in
the expansion of the coupling $V_k$ in powers of $\epsilon_k$
\cite{moreterms}: $V_k\approx V_{k_F}+\epsilon_k(dV/d\epsilon_k)_{k_F}$. We
rewrite this approximation as $V_k= V +\pi \rho q V \epsilon_k$, where
$\rho\equiv1/2D$, to define the coupling $V$ and the Fano parameter
$q$ \cite{Fa61:1866}. Following NRG tradition, we introduce the shortand
$f_0\equiv\sum_k c_k/\sqrt{N}$. For future reference, we note that, in
this notation, the (particle-hole) {\em symmetric} ($q=K=2\ed+U=0$)
Anderson Hamiltonian reads
\begin{equation}
  \label{eq:12}
H_{0}^S= \sum_k\epsilon_k c_k^\dagger c_k + V(f_0^\dagger\cd +\hc)
-\frac{U}2(\ndu-\ndd)^2.
\end{equation}

The conductance is more easily computed on a basis constituted of
$f_0$ and $N-1$ other conduction operators $a_p=\sum_k \alpha_{pk}
c_k$, where $p=2\pi n_p/L$ [$n_p=-N/2+1,\ldots,N/2-1$], such that
$\{f_0^\dagger, a_p\}=0$, and that $\{a_p^\dagger,
a_{p'}\}=\delta_{pp'}$ \cite{apbasis}. With the shorthand $f_1= \sum_p
a_p/\sqrt{N-1}$, the model Hamiltonian becomes
\begin{equation}\label{eq:6}
 \ha =\sum_p\tilde\epsilon_p a_p^\dagger a_p 
 +Kf_0^\dagger f_0 +(t_0\fone^\dagger \dq
 + Vf_0^\dagger\cd +\hc) +H_d,
\end{equation}
where $\nq\equiv\sqrt{1+(\pi\rho Vq)^2}$; $\nq d_q\equiv f_0+
\pi\rho\,q\,V\cd$; $t_0\equiv D\nq/\sqrt3$; and the energies
$\tilde\epsilon_k$ are the conduction energies $\epsilon_k$ phase
shifted by $\pi/2$, \ie~$\tilde\epsilon_{k}\equiv
\epsilon_{k}-\Delta/2$.

If $K\to\infty$, the scattering potential decouples $f_0$ from the
other states, freezes its occupation at $f_0^\dagger f_0=0$, and
forces the current through the dot. The condition $2\ed+U=0$
reduces $\ha$ to the {\em symmetric-SET} Hamiltonian
\begin{equation}\label{eq:6a}
H^S_{SET}= \sum_p\tilde\epsilon_p a_p^\dagger a_p + V_q(f_1^\dagger\cd +\hc)
-\frac{U}2(\ndu-\ndd)^2,
\end{equation}
where $V_q\equiv\pi\,q\,V/(2\sqrt3)$. Given the analogous definitions
$f_0\sim\sum_k c_k$ and $f_1\sim\sum_pa_p$, we see that only the phase shifted
energies $\tilde\epsilon_k=\epsilon_k-\Delta/2$ distinguish
$H^S_{SET}$ from $H_0^S$.

The first term within parentheses on the right-hand side of
  Eq.~(\ref{eq:6}), which couples the conduction states $f_1\sim\sum
  a_p$ to $d_q\sim f_0+ \pi\rho\,q\, V\cd$, is the mathematical
  expression of the two conduction paths in Fig.~\ref{fig:1}, one of
  which runs through the central wire-orbital $f_0$, and the other,
  with relative amplitude $\pi\rho\,q\,V$, through the dot orbital
  $\cd$. To flow from the left to the right side of the wire, the
  current must traverse $d_q$. Accordingly, the Linear-Response Theory
  \cite{Gre58:585} shows that $\rho_q$, the spectral density
  for the operator $d_q$, controls the zero-bias conductance:
\begin{equation}
  \label{eq:5}
{G_q(T)}=\gc{\nq^2}
\int_{-D}^{D}\frac{\rho_{q}(\epsilon,T)}{\rho} \left[-\frac{\partial
    f(\epsilon)}{\partial\epsilon}\right]\,\upd\epsilon.
\end{equation}
Here $f(\epsilon)$ is the Fermi function, and
\begin{equation}
  \label{eq:7}
\rho_q(\epsilon,T)= \frac1{\mathcal{Z}}\sum_{mn}\frac{e^{-\beta
    E_m}}{f(\epsilon_{mn})}|\matel{m}{d_q^\dagger}{n}|^2
\delta(\epsilon_{mn}-\epsilon),
\end{equation}
where $\mathcal{Z}$ is the partition function, and
$\epsilon_{mn}\equiv E_m-E_n$.

{\em Analysis.}  We are chiefly interested in the {\em Kondo regime},
the set of temperatures and model parameters favoring unitary dot
occupation, \ie~such that the energies $|\ed|$ ($\ed+U$) to remove the
dot electron (add a second electron) dwarf the energy $k_BT$ and
dot-level width $\Gamma=\pi\rho V^2$. The Kondo temperature $T_K$ then
sets the energy scale.

At high temperatures, $T\gg T_K$, the Hamiltonian~(\ref{eq:6}) lies
close to an unstable {\em local-moment} fixed point (LM), which
comprises a noninteracting spin-1/2 variable (the dot spin)
decoupled from a conduction Hamiltonian \cite{KWW80:1044}
\begin{equation}
  \label{eq:14}
H^*_{LM} = \sum_k \epsilon_kc_k^\dagger c_k+ K_{LM}f_0^\dagger f_0,
\end{equation}
where the effective scattering potential $K_{LM}$ depends on the model
parameters. For $\ha=H_0^S$ ($H_{SET}^S$), in particular,
$K_{LM}=0$ ($K_{LM}\to\infty$).

Diagonalization of the quadratic form~(\ref{eq:14}) yields $N$
eigenoperators $g_k$ and eigenvalues $\varepsilon_k=\epsilon_k - \delta_{LM}\Delta/\pi$:
\begin{equation}
  \label{eq:15}
  H^*_{LM} = \sum_k \varepsilon_k g_k^\dagger g_k.
\end{equation}
Near the Fermi level, the phase shifts $\delta_{LM}$ are uniform \cite{KWW80:1044}.

In analogy with the definitions $f_0\sim \sum_k c_k$ and $f_1
\sim\sum_k\epsilon_k c_k$, we can define the mutually orthogonal
combinations of eigenoperators $\phi_0=\sum_k g_k/\sqrt{N}$ and
$\phi_1=\sqrt{3/N}\sum_k (\epsilon_k/D)g_k$. For any pair of
  constants $\alpha_0$ and $\alpha_1$,
\begin{equation}
  \alpha_0 f_0 +\alpha_1 f_1 = \beta_0 \phi_0 + \beta_1\phi_1,
    \label{eq:18}
\end{equation}
where $\beta_0$ and $\beta_1$ are linear combinations of $\alpha_0$
and $\alpha_1$, with coefficients fixed by $\delta_{LM}$.

Table~\ref{tab:0} collects results for the symmetric
  [Eq.~(\ref{eq:12})] and the symmetric-SET [Eq.~(\ref{eq:6a})]
  Hamiltonians. While columns 3-5 follow from the definitions of
  $g_k$, $\phi_0$ and $\phi_1$, columns 5 and 6 require
  explanations. For $\ha=H_0^S$ (\ie\ $q=K=2\ed+U=0$), the operator $d_q$ reduces to
  $f_0=\phi_0$. Its spectral density
  $\rho_q\equiv\rho_0^S$ is therefore equal to $\rho_{\phi_0}$.
\begin{table}[th]
    \caption[sym and SET]{\label{tab:0}Properties of the symmetric
      Hamiltonians}
    \centering
\begin{tabular}{ccccccc}
      \hline\hline\\[-2ex]
      $\ha$ &  $\delta_{LM}$ & $g_k$ & $\phi_0$ & $D\phi_1$
      &$\rho_{\phi_0}$&$\displaystyle\frac{\rho_{\phi_1}}{\nq^2}$\\
      \\[-2ex]
      \hline
      \\[-1.5ex]
      $H_0^S$ & 0 & $c_k$ & $f_0$ & $Df_1$ & $\rho_0^S$ & \\
      $H_{SET}^S$ & $\pi/2$ & $a_k$ &  $f_1$  &$\sqrt{\frac{3}{N}}
        \sum_p\tilde\epsilon_pa_p$& &$\rho_{SET}^S$\\[0.5ex]
      \hline\hline
    \end{tabular}
\end{table}

For $\ha=H_{SET}^S$, $d_q$ reduces
  $-\pi\rho\,q\,V\,c_d/\nq$. To relate its spectral density to
  $\rho_{\phi_1}$, choose two eigenstates $\ket m$, $\ket n$ of
  $H_{SET}^S$ with $E_m,E_n\approx k_BT\ll D$. In the identity
  \begin{equation}\label{eq:commut}
    \matel{m}{\sum_p\frac{[a_p,H_{SET}^S]}{\sqrt N}}{n}=
    \frac{D\matel{m}{\phi_1}{n}}{\sqrt3}+{V_q}\matel{m}{\cd}{n},
  \end{equation}
  the left-hand side is then much smaller than each
  matrix element on the right. This shows that
  $\matel{m}{\phi_1}{n}\approx-\pi\rho\,q\,V\matel{m}{\cd}{n}$, a result
  equivalent to $\matel{m}{\phi_1}{n}= -\nq\matel{m}{\dq}{n}$, and
  hence to the last column in Table~\ref{tab:0}.

As the system is cooled, the wire electrons screen the dot moment, and
$\ha$ crosses over from the LM to a stable {\em frozen-level} fixed
point (FL). The latter is but the conduction band resulting from
letting $K_{LM}\to 1/(\pi^2\rho^2K_{LM})$ in
Eq.~(\ref{eq:14}). Diagonalization of the FL Hamiltonian yields $N$
eigenvalues $\overline\varepsilon_k=\epsilon_k - \delta\Delta/\pi$,
where, in conformity with Friedel's sum rule \cite{La66:516}, the
phase shift $\delta=\delta_{LM}-\pi/2$.

\begin{table}[th]
  \caption{\label{tab:1}\sf Conductance at the two fixed points }
  \centering
    \begin{tabular}{cccccc}
      \hline\hline\\[-2ex] Fixed& Phase & $G_q$ & $G_0^S$ &
      $G_{SET}^S$\\ point &shift&&$(\delta=\pi/2)$&$(\delta=0)$
      \\[0.25ex]
      \hline\\[-1.5ex] LM &
      $\delta+\pi/2$ & $\gc\sin^2\delta$ & $\gc$ & 0 \\ FL & $\delta$
      & $\gc\cos^2\delta$ & 0 & $\gc$ \\[0.5ex] \hline\hline
    \end{tabular}
\end{table}
The fixed-point physical properties are independent of temperature and
energy. In particular, the LM and FL conductances are trigonometric
functions of the phase shift. The expressions, which result from an
extension of Langreth's argument \cite{La66:516}, are recorded in
Table~\ref{tab:1}.

The effective antiferromagnetic interaction
$H_{J}={J\bm{S}\!\bm{\cdot}\!\bm{\sigma}_{\mu\nu}\phi_{0\mu}\phi_{0\nu}}$
\cite{SW66:491,KWW80:1003,KWW80:1044}, between the dot spin $\bm{S}$
and the spin of the localized orbital $\phi_0$ drives the Hamiltonian
from the LM to the FL.  The NRG~trajectory is universal: scaled by
$k_BT_K$, the eigenvalues of $\ha$ are universal, and so are the
corresponding eigenstates on the basis of the $\{g_k\}$
\cite{universalprops,Wi75:773,KWW80:1044}. The spectral densities
$\rho_{\phi_0}$ and $\rho_{\phi_1}$ in Table \ref{tab:0} are therefore
universal functions of the ratios $T/T_K$ and $\epsilon/k_BT_K$ . To
highlight these findings, we define the scaled energy $\esc\equiv
\epsilon/k_BT_K$ and temperature $\tsc\equiv T/T_K$.

Next, we turn to the asymmetric Kondo-domain Hamiltonians. To
relate the conductance $G_q$ to the universal function $G_{SET}^S$, we
add to $\ha$ an infinitesimal harmonic perturbation, frequency
$\omega$, coupling the spectrum of $\ha$ to an auxiliary orbital $\dd$
at the Fermi level:
\begin{equation}\label{eq:10}
  \heta = \eta\,\ddd d_q\,e^{-i\omega t}+\hc.
\end{equation}

The golden rule shows that the spectral density $\rho_q$ is
the response function for the thermally averaged transition rate
$j{\>}_{\dd\to A}$ induced by $\heta$:
\begin{equation}
  \label{eq:4}
  \langle j{\>}_{\dd\to A}(\omega)\rangle_T =
  (\pi\eta^2/\hbar)\rho_q(\hbar\omega, T).
\end{equation}

Consider, then, the perturbative effects of $\heta$ upon the Kondo
crossover \cite{KWW80:1003}.  Close to a (Fermi-liquid) fixed point
$H^*$, that is, for $H=H^*+\delta H$, one can always construct an
effective Hamiltonian $H_{eff}$ that reproduces the spectrum of $H$ to
linear order in $\delta H$. Here, to follow a pedestrian route, we
subject the sum $\ha+\heta$ to the Schrieffer-Wolff transformation:
$H_A^{eff}+\heta^{eff}\equiv e^S(\ha+\heta)e^{-S}$ \cite{SW66:491}.
Besides substituting the spin-spin interaction $H_J$ for the dot-wire
coupling $H_{wd}$, this casts Eq.~(\ref{eq:10}) in the form
\begin{equation}\label{eq:20}
  \heta^{eff} = (\eta/\nq)\ddd (\alpha_d c_d+ \alpha_0 f_0 +
  \alpha_1 f_1)e^{-i\omega t}+\hc,
\end{equation}
where $\alpha_d$, $\alpha_0$, and $\alpha_1$ depend on $V$,
$\ed$, $U$, and $K$.

The unperturbed effective Hamiltonian $\ha^{eff}$ commutes with
$n_d$. Of its eigenstates, only those with $n_d=1$ are energetically
accessible at the LM. It is safe to disregard the perturbation
proportional to $\alpha_d$ in Eq.~(\ref{eq:20}), which couples them to
the subspaces $n_d=0$ and $n_d=2$.

The other two perturbations conserve $n_d$ and are important even at
very low energies. It is convenient to project them upon $\phi_0$ and
$\phi_1$, because the spectral densities for $f_0$ and $f_1$ are
phase-shift dependent. Substitution of Eq.~(\ref{eq:18}) for
  $\alpha_0f_0+\alpha_1f_1$ brings Eq.~(\ref{eq:20}) to the form
\begin{equation}\label{eq:11}
  \heta^{eff}= (\eta/\nq) \ddd(\beta_0\,\phi_0 + \beta_1\,\phi_1) +\hc.
\end{equation}

In the (LM to FL) crossover, just as the Kondo Hamiltonian
$\ha^{eff}=e^S\ha e^{-S}$ is equivalent to $\ha$ and the
diagonalization of $\ha^{eff}$ yields the physical properties for
$\ha$, the effective perturbation $\heta^{eff}$ is equivalent to
$\heta$ and application of the golden rule to $\heta^{eff}$ yields
the transition rate induced by $\heta$. Out of the terms then
resulting from Eq.~(\ref{eq:11}), only those proportional to
$|\matel{m}{\phi_0^\dagger}{n}|^2$ and
$|\matel{m}{\phi_1^\dagger}{n}|^2$ contribute to $G_q$
\cite{particle-hole}. The cross terms disregarded, the last
  two columns in Table \ref{tab:0} lead to
\begin{equation}
  \langle j{\>}_{\dd\to A}\rangle_T =
  \frac{\pi\eta^2}{\hbar}(\frac{\beta_0^2}{\nq^2}\rho_0^S+\beta_1^2\,\rho_{SET}^S).
\end{equation}
Comparison with Eq.~(\ref{eq:4}) then shows that 
\begin{equation}
  \label{eq:8}
\rhoq = (\beta_0^2/\nq^2)\rho_{0}^S(\esc,\tsc)+\beta_1^2\,\rho_{SET}^S(\esc,\tsc),
\end{equation}
and Eq.~(\ref{eq:5}) yields an expression for the conductance:
\begin{equation}\label{eq:13}
  G_q(\tsc)= \beta_0^2\,G_{0}^S(\tsc) +\nq^2\beta_1^2\,G_{SET}^S(\tsc).
\end{equation}

Substitution of the expressions in the LM line of Table~\ref{tab:1} for
$G_q$, $G_{0}^S$, and $G_{SET}^S$ shows that
$\beta_0^2=\sin^2\delta$, while the expressions in the FL line show that
$\nq^2\beta_1^2=\cos^2\delta$. With this, Eq.~(\ref{eq:13}) becomes
\begin{equation}
  G_q(\tsc)=G_0^S(\tsc)\,\sin^2\delta +G_{SET}^S(\tsc)\,\cos^2\delta,
\end{equation}
and to complete the derivation of Eq.~(\ref{eq:2}) we only have to recall
that $G_0^S=\gc-G_{SET}^S$.

\begin{table}[htb]
  \caption{\sf Parameters for the five runs in Fig.~\ref{fig:2}b. $U=0.20\,D$.}
  \label{tab:2}
  \centering
  \begin{tabular}{cccccc}
    \hline\hline\\[-2ex]
    $-\ed/D$ & $V/D$ & $-K/D$ & $q$&$\delta/\pi$&$k_BT_K/D$\\[0.25ex]
    \hline\\[-1.5ex]
    0.10 & 3.6$\times10^{-4}$&100.0&100&0.00&1.3$\times10^{-5}$ \\
    0.17 & 0.021 & 0.315 & 1& 0.22 &1.1$\times10^{-5}$\\
    0.10 & 0.021 & 0.315 & 1&0.25& 1.1$\times10^{-9}$\\
    0.022 & 0.021 & 0.315 & 1&0.28& 1.1$\times10^{-5}$\\
    0.10 & 0.056 & 0.000 & 0&0.5 & 1.4$\times10^{-5}$\\[0.5ex]
    \hline\hline
  \end{tabular}
\end{table}

{\em Discussion.}  The phase shift $\delta$ and Kondo temperature
$T_K$ in Eq.~(\ref{eq:2}) depend on the model parameters. Almost
invariably, they have to be computed numerically, because the
perturbative expressions for $\delta$ and $T_K$ are accurate only in
corners of the parametrical space
\cite{Wi75:773,KWW80:1044,AFL83:331,TW83:453,hewson93}. Exceptions are
the particle-hole symmetric Hamiltonians $H_0^S$ [Eq.~(\ref{eq:12})]~and
$H_{SET}^S$ [Eq.~(\ref{eq:6a})], for which $\delta=\pi/2$ and $\delta=0$, so that
Eq.~(\ref{eq:2}) reduces to
$G_q(\tsc)=\gc-G_{SET}^S(\tsc)=G_0^S(\tsc)$ and 
$G_q(\tsc)=G_{SET}^S(\tsc)$, depicted in the right and left panels of
Fig.~\ref{fig:2}b, respectively. 

In other regions of the Kondo domain, $\delta$ lies between $-\pi/2$
and $\pi/2$. In the Kondo crossover, the conductance on the left-hand
side of Eq.~(\ref{eq:2}) changes by less than $\gc$, from
$\gc\sin^2\delta$ to $\gc\cos^2\delta$.  The rise or decay of $G_q(T)$
is centered at $\gc/2$ and proportional to $\gu-\gc/2$.

The central panel of Fig~\ref{fig:2}b shows examples. The solid lines
are Eq.~(\ref{eq:2}) with $\delta$ (Table~\ref{tab:2}) extracted from
the FL single-particle eigenvalues in the NRG diagonalization of the
pertinent Hamiltonian $\ha$; and $T_K$ (Table~\ref{tab:2}), from a fit
of the universal magnetic susceptibility \cite{KWW80:1003,TW83:453} to
the susceptibility computed for the same Hamiltonian. No adjustable
parameter is therefore involved in the excellent agreement between
Eq.~(\ref{eq:2}) and the NRG data for the conductance. We have applied
the same procedure to more than 100 NRG runs sampling the Kondo
domain; in each one, the agreement was equally good.

In brief, the exact universal mapping~(\ref{eq:2}) interpolates
between $G(T\gg T_K)$ and $G(T\ll T_K)$. While the two extremes,
described by single-particle Hamiltonians, are accessible to a variety
of techniques\textemdash \eg~the Landauer-Buttiker formula or
scattering-matrix analyses \cite{Bu86:1761}\textemdash, the
interpolation covers the temperature range beyond the reach of simple
analyses.
\begin{figure}
\onefigure[width=1.0\columnwidth]{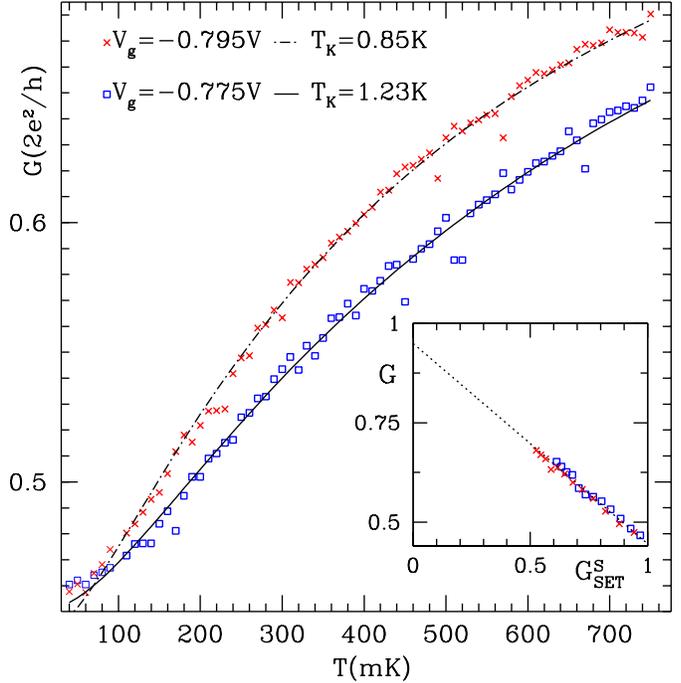}
\caption{Comparison with experiment \cite{SAK+05:066801}.  The crosses
  and squares are the conductances $G$ at the indicated gate
  potentials $V_g$; the temperatures measured below $100\text{mK}$
  carry large uncertainties \cite{SAK+05:066801}. The dash-dotted
  and solid lines depict Eq.~(\ref{eq:3}). The inset shows that, for
  each $V_g$, the appropriate $T_K$ straightens the plot of $G$
  vs.\ $\gs$ and yields Eq.~(\ref{eq:3}).}
  \label{fig:3}
\end{figure}

{\em Comparison with experiment.} To allow for the inevitable
background currents, we exploit the linearity in
Eq.~(\ref{eq:2}). Given a set of $N$ experimental pairs $\{G_i,
T_i\}$, from a trial Kondo temperature $T_K^*$ we generate the
dimensionless temperatures $\tsc_i = T_i/T_K^*$
($i=1,\ldots,N$). Next, we invert the function $\gs(\tsc)$ to
determine the universal conductance $\gs_i$ for each $\tsc_i$. If the
plot of $G_i$ vs.\ $\gs_i$ is straight, $T_K^*$ is the Kondo
temperature. If it is not, we iterate. Visual inspection (numerical
evaluation of the curvature) determines $T_K$ within $10\%$ error (to
an accuracy limited only by the experimental dispersion).

As an example, the inset of Fig.~\ref{fig:3} treats the conductances
in Fig.~3b of Ref.~\onlinecite{SAK+05:066801}, measured with the gate
potentials $V_g=-795\,\text{mV}$ (squares) and $-775\,\text{mV}$
(crosses), and yields the Kondo temperatures $T_K=850\,\text{mK}$ and
$1230\,\text{mK}$, as well as the linear relation
\begin{equation}
\label{eq:3}
  G(\tsc) =1.9e^2/h-0.5\gu(\tsc).
\end{equation}
For $T\gg T_K$, $\gu(\tsc)\to0$, so that $G\to1.9e^2/h$,
close to the measured off-resonance conductance ($1.8e^2/h$)
\cite{SAK+05:066801}.

The agreement with the solid and the dash-dotted lines representing
Eq.~(\ref{eq:3}) in Fig.~\ref{fig:3} shows that the data are in the
Kondo regime and the contact resistance is negligible. There is,
however, a background current: at $\tsc=1$, Eq.~(\ref{eq:3}) yields
$G(\tsc=1)=1.4e^2/h$; the excess $G_b=0.4e^2/h$ over $G(\tsc)=e^2/h$,
predicted by Eq.~(\ref{eq:2}) is the background conductance
\cite{SAK+05:066801}. If we define $\tilde G \equiv G-G_b$, then
Eq.~(\ref{eq:3}) takes the universal form $\tilde G - e^2/h= -0.5(\gu
-e^2/h)$, from which we find the shift: $\cos2\delta=-0.5$
($\delta\approx\pi/3)$.

In conclusion, we have shown that, in the Kondo regime, measured from
$\gc/2$ and scaled by $\cos2\delta$, the conductance of side-coupled
devices is a universal function of $T/T_K$. Our application to
experimental data identified inequivocally the measured thermal
dependences with Kondo screening; detected a background current; and
determined the ground-state phase shift.

\begin{acknowledgments}
We are grateful to Drs.~Alvaro Ferraz, Vivaldo Campo,
V.~V.~Ponomarenko, and R.~Pepino for stimulating discussions; to
Prof.~Shingo Katsumoto for the data in Fig.~\ref{fig:3}. The CNPq,
FAPESP (01/14974-0; 04/08928-3), and IBEM supported this work.
\end{acknowledgments}


\begin{thebibliography}{10}
\bibitem{even}
Given the inversion symmetry of Fig.~\ref{fig:1}, we define conduction states
  $c_k$ ($b_k$) with even (odd) parity. Decoupled from the dot, the $b_k$'s
  need not appear in $H_A$.
\bibitem{moreterms}
Neither our analysis nor its result [Eq~(\ref{eq:2})] depend on the number of
  terms in the expansion of $V_k$.
\bibitem{universalprops}
Projected on the $\{g_k\}$ basis, $\ha^{eff}/k_BT_K$ has universal
  eigenvalues and eigenvectors. Thermally averaged expectation values involving
  only the operators $\phi_0\sim \sum_k g_k$ and $\phi_1\sim\sum_k\varepsilon_k
  g_k$ are therefore universal functions.
\bibitem{particle-hole}
The contribution to $\rho_q$ of the cross terms,
  $\rho_{01}(\epsilon)\equiv\sum_{mn}e^{-\beta
  E_m}/f(\epsilon)\matel{m}{\phi_0^\dagger}{n}
  \matel{n}{\phi_1}{m}\delta(\epsilon-\epsilon_{mn})+c.~c.$, is universal
  \cite{universalprops}. We can therefore choose $\ha=H_0^S$ to compute it.
  Given that $H_0^S$ is invariant under particle-hole transformations, while
  $\rho_{01}(\epsilon)\to-\rho_{01}(-\epsilon)$, we see that
  $\rho_{01}(\epsilon)$ is an odd function of the energy; substituted for
  $\rho_q(\epsilon)$ in Eq.~(\ref{eq:5}), it makes no contribution to $G_q$.
\end{thebibliography}

\end{document}